\documentclass[journal]{IEEEtran}

\usepackage{cite}
\usepackage{amsmath,amssymb,amsfonts}
\usepackage{algorithmic}
\usepackage[dvipsnames]{xcolor}
\usepackage{colortbl}
\usepackage{pgfplots}
\usepackage{verbatim}
\pgfplotsset{width=7cm,compat=1.8}
\usepgfplotslibrary{fillbetween}
\usepackage{xcolor}
\usepackage{tikz}
\usepackage{tikzscale}
\usepackage{pgfplots}
\usetikzlibrary{patterns}
\pgfplotsset{every tick/.style={black,}}
\usepackage{algorithm,algorithmic}
\usepackage{subcaption}
\usepackage{caption}
\usepackage{amsmath}
\usepackage{booktabs}
\usepackage{multirow}
\usepackage{url}
\usepackage{pgfplotstable}
\usepackage[normalem]{ulem}
\usepackage{stackengine}
\usepackage{arydshln}

\newcommand\xrowht[2][0]{\addstackgap[.5\dimexpr#2\relax]{\vphantom{#1}}}

\pgfplotsset{compat=1.11,
        /pgfplots/ybar legend/.style={
        /pgfplots/legend image code/.code={%
        \draw[##1,/tikz/.cd,bar width=8pt,yshift=-0.2em,bar shift=0pt]
                plot coordinates {(0cm,0.8em)};},
},
}
\newcolumntype{g}{>{\columncolor{blue!10}}c}

\begin{document}

\title{
Neural Network-based OFDM Receiver for Resource Constrained IoT Devices 
{\footnotesize \textsuperscript{}}
\thanks{*Hai Cheng, Mauro Belgiovine, Yanyu Li, and Haoqing Li have equally contributed to this paper.}
}

\author{
\IEEEauthorblockN{
Nasim Soltani, Hai Cheng*, Mauro Belgiovine*, Yanyu Li*, Haoqing Li*, Bahar Azari, Salvatore D'Oro, Tales Imbiriba, Tommaso Melodia, Pau Closas, Yanzhi Wang, Deniz Erdogmus, and Kaushik Chowdhury
}\\
\IEEEauthorblockA{
Electrical and Computer Engineering Department, Northeastern University, Boston, MA
}\\
\IEEEauthorblockA{\textit{\{soltani.n, cheng.hai, belgiovine.m, li.yanyu, li.haoq\}@northeastern.edu, azari@ece.neu.edu, \{s.doro, talesim, t.melodia, pau.closas, yanz.wang\}@northeastern.edu, \{erdogmus, krc\}@ece.neu.edu}}
}

\markboth{Accepted in IEEE IoT Magazine, May 2022}%
{}

\maketitle

\begin{abstract}
Orthogonal Frequency Division Multiplexing (OFDM)-based waveforms are used for communication links in many current and emerging Internet of Things (IoT) applications, including the latest WiFi standards. For such OFDM-based transceivers, many core physical layer functions related to channel estimation, demapping, and decoding are implemented for specific choices of channel types and modulation schemes, among others. To decouple hard-wired choices from the receiver chain and thereby enhance the flexibility of IoT deployment in many novel scenarios without changing the underlying hardware, we explore a novel, modular Machine Learning (ML)-based receiver chain design. Here, ML blocks replace the individual processing blocks of an OFDM receiver, and we specifically describe this swapping for the legacy channel estimation, symbol demapping, and decoding blocks with Neural Networks (NNs). A unique aspect of this modular design is providing flexible allocation of processing functions to the legacy or ML blocks, allowing them to interchangeably coexist. Furthermore, we study the implementation cost-benefits of the proposed NNs in resource-constrained IoT devices through pruning and quantization, as well as emulation of these compressed NNs within Field Programmable Gate Arrays (FPGAs). Our evaluations demonstrate that the proposed modular NN-based receiver improves bit error rate of the traditional non-ML receiver by averagely 61\% and 10\% for the simulated and over-the-air datasets, respectively. We further show complexity-performance tradeoffs by presenting computational complexity comparisons between the traditional algorithms and the proposed compressed NNs.
\end{abstract}


\section{Introduction}\label{sec:intro}

The Internet of Things (IoT) paradigm will enable exciting applications, such as remote surgery, autonomous cars, augmented/virtual reality, all of which demand faster processing, higher data rates, and more reliable communications beyond what is realizable today. To meet such demands, most modern WiFi standards deployed in commercial consumer IoT devices today, including IEEE 802.11a/g/n/ac and other emerging standards use Orthogonal Frequency Division Multiplexing (OFDM).

\begin{figure}[t!!!]
    \centering
    \includegraphics[width=\linewidth]{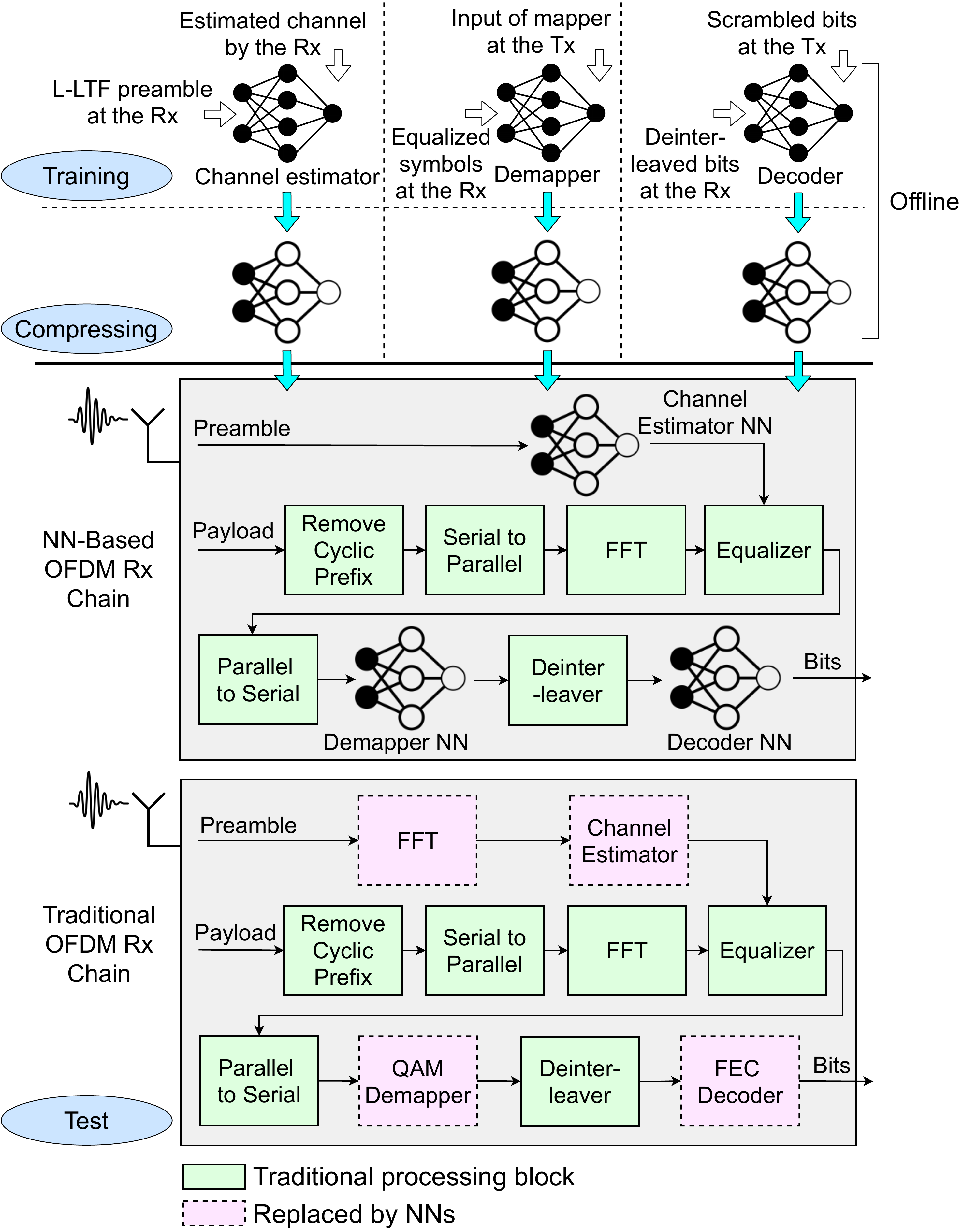}
    \caption{An overview of the proposed Neural Network (NN)-based receiver. The channel estimator, demapper, and error corrector (decoder) blocks are substituted with individual NNs. The NNs are trained and compressed offline, for online FPGA deployment.}
    \label{fig:overview}
\end{figure}

The typical OFDM receiver consists of several signal processing blocks that detect and synchronize the packet, estimate the channel and equalize the payload to overcome channel-induced distortions, and finally extract useful bit representations through demapping and error correction (decoding). An overview of the traditional processing blocks is shown in Fig.~\ref{fig:overview} (bottom).
Indeed, hand-engineered processing steps (e.g., using custom-designed packet preambles or fixed modulation schemes) offer limited opportunities for on-the-fly adaptation. Moreover, the wireless environment is too complex to be modeled accurately, and constraining the choice of the processing blocks to only one of several candidates may lower the performance.
On the other hand, Neural Networks (NNs) provide an adaptable and noise-resilient solution for many physical layer processing tasks, such as modulation classification~\cite{soltani2019spectrum} and RF fingerprinting~\cite{augmentation,nasim-uav}, that improve the performance of their traditional counterparts. Similarly, in the domain of receiver design, NNs can offer a closed-form and flexible solution by learning to imitate previous channel estimations, symbol demappings, and decodings, instead of explicitly realizing the mathematical form of these processing blocks.

\noindent $\bullet$ \textbf{Real-time Computation.} The flexibility and performance improvement arising from including ML blocks within the receiver processing chain comes at a computational cost. Although NN computations are simple multiplications and additions, NNs are overall compute-intensive, and thus deploying them might cause delays that impact time-sensitive applications. In the wireless receiver, delays even as small as a few microseconds might prevent the processing chain to remain synchronized with the streaming data, which causes incorrect decoding. Thus, methods to achieve real-time NN computations are critical.


\noindent $\bullet$ \textbf{Proposed Modular Design.}
While a single NN that captures the entire receiver processing chain simplifies the OFDM receiver, the routing of data, and NN inference execution, there are several drawbacks of such a monolithic design. First, this becomes a `black-box' approach that yields no insights on the performance of the intermediate functional steps. Thus, if say the \textit{demapper} is under-performing in a given situation, there is no way for the designer to know this. Second, the receiver may be frequently deployed in new wireless environments, which necessitates re-training of the entire NN with massive volumes of data each time. For example, introducing a new modulation scheme or coding rate to the original waveform renders the entire prior training inadmissible.
These considerations motivate us to pursue a model-driven design, with the goal of maintaining full compatibility with the classical processing chain. Thus, any individual classical processing block could be swapped with its ML block counterpart in a way that is transparent to the rest of the receiver chain. This fully modular approach distinguishes our work from other recent work that use NN-based solutions for cyclic prefix free~\cite{CP-free}, DFT free~\cite{deep-waveform} or pilotless communications~\cite{Pilotless} in OFDM systems. 

In this paper, we propose an NN-based end-to-end OFDM receiver. Our scheme is composed of channel estimator, demapper, and error corrector (decoder) NNs with totally $\sim$3.1M parameters, cascaded to build a complete receiver, as shown in Fig.~\ref{fig:overview} (top). Using both simulated and Over-The-Air (OTA) datasets, we show that these NNs perform better than their non-ML counterparts. Moreover, we propose Block Column Row (BCR) pruning and Mixed Scheme Quantization (MSQ) to compress these models without accuracy loss for Field Programmable Gate Arrays (FPGA) deployment, and present FPGA results. We further calculate computational complexity in terms of floating-point operations (FLOPs) for the traditional and NN algorithms and provide comparisons of the two. Our prototype design can be implemented in small form-factor FPGAs that may be present in IoT devices, or it can be used to design a custom chip for NN-based OFDM receivers for specific IoT applications.



\section{Related Work}\label{sec:related}

\begin{figure}[t!!!]
    \centering
    \includegraphics[width=\linewidth]{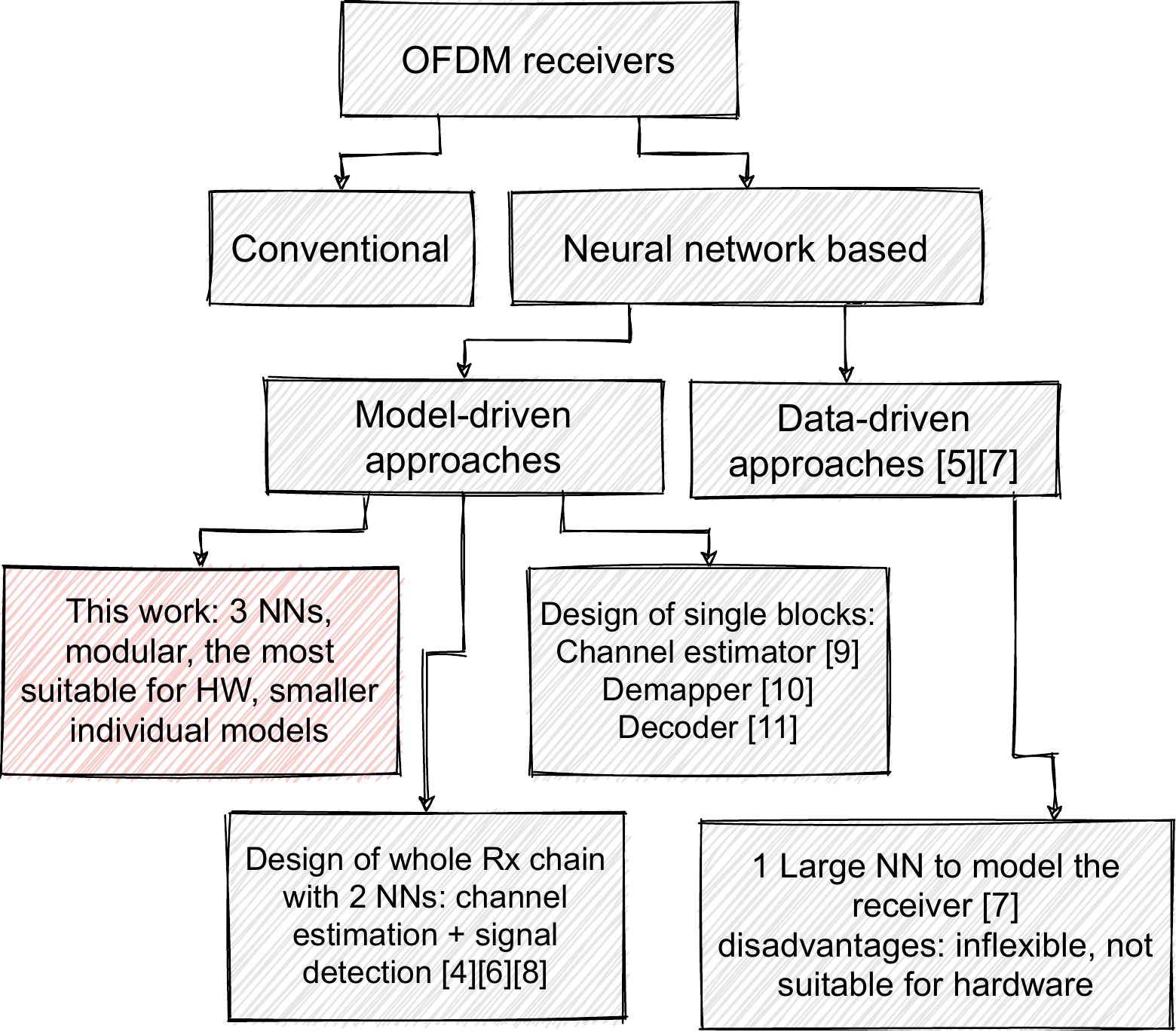}
    \caption{Related work summary for OFDM receivers.}
    \label{fig:related}
\end{figure}

We categorize the previous work on modeling the OFDM receiver chain into (i) data-driven approaches, where the NN models are developed using data, without domain knowledge being involved, and (ii) model-driven approaches, where domain knowledge is used in the design of NN-based receivers. In~\cite{data-driven-letter}, a data-driven approach is proposed that directly predicts the transmitted symbols without explicit channel estimation, by a single deep NN. This NN is trained to minimize the difference between its output and the transmitted data.
In~\cite{deep-waveform}, the authors eliminate the DFT in the OFDM receiver using a complex-valued deep NN. They train their channel equalizer based on a frozen pre-trained basic receiver.

In contrast, in model-driven design, domain knowledge is adopted to design separate NNs for each function in the receiver chain~\cite{gao2018comnet}. \cite{CP-free} proposes an AI-aided OFDM chain for a cyclic prefix-free system, where channel estimation and signal detection are done by separate NNs. The authors integrate the orthogonal approximate message passing algorithm with the signal detection NN. 
The authors in~\cite{Pilotless} propose end-to-end learning with the purpose of eliminating orthogonal pilots in OFDM symbols by jointly optimizing parts of the transmitter and the receiver.

There are additional work on replacing individual blocks in the traditional receiver with NNs. For example, \cite{mauro-channelEst} proposes a multi-layer perceptron NN for estimating the channel in massive MIMO systems for 5G and beyond.
Authors in~\cite{demapper} propose a soft-demapper using a fully-connected NN. Aside from the OFDM receiver context, authors in~\cite{decoder} propose a recurrent neural network to decode different coding schemes, including the convolutional codes in the Viterbi algorithm~\cite{viterbi1967error}.

A summary of the aforementioned categories is shown in Fig.~\ref{fig:related}. Despite the large body of literature that propose NNs for performing different tasks in OFDM receivers, to the best of our knowledge, prior work do not propose a fully-modular NN design for OFDM receivers, which is customized for implementation on resource-constrained hardware.


\section{Model-driven NN-Based OFDM Receiver}\label{sec:proposed}
In this section, we describe channel estimation, demapping, and decoding tasks in the OFDM receiver, through both the classical and NN methods.
These NNs are individually trained and then cascaded to build the complete receiver. Finally, we describe our pruning and quantization methods to compress these NNs for FPGA implementation.

\subsection{Channel Estimation}\label{sec:proposed-channel}

\noindent $\bullet$ \textbf{Classical Approach.} 
After packet detection, the OFDM wireless receiver estimates the Channel State Information (CSI) to compensate for channel-induced distortions in the received signal. This estimation is done in the frequency domain, for different OFDM formats in WiFi 802.11a/g/n/ac, by leveraging the Legacy Long Training Field (L-LTF) in the preamble.

The L-LTF consists of 2 identical OFDM symbols containing pilot information, known at the receiver. The Least Square (LS) channel estimation is used to estimate the channel coefficients of each sub-carrier. In the traditional receiver, after OFDM demodulation,
LS method divides the received pilot sequence element-wise by the known pilot sequence. The final CSI is the average of the two different CSI vectors obtained by performing the LS estimation on the two OFDM symbols in the L-LTF.

\noindent $\bullet$ \textbf{Proposed NN Architecture.}
In contrast to the classical channel estimator whose input is frequency-domain L-LTF, our proposed channel estimator NN processes the time-domain L-LTF.
This obviates the need for the FFT operation during demodulation, as shown in Fig.~\ref{fig:overview}, which is beneficial as FFT becomes expensive for large number of subcarriers.
As the estimated channel is calculated independently for each sub-carrier, we use fully-connected layers to design the channel estimator NN. Our NN consists of two Multi-Layer Perceptron (MLP) architectures that estimate the real and imaginary parts of CSI independently. Each MLP has an input size of 160 with 2 hidden layers, of 512 and 256 neurons, respectively. The hidden layers have ReLU activation. Without loss of generality, we use WiFi 802.11a-compliant waveform that has 52 channel coefficients. Thus, the channel estimator output layer also is of size 52, with linear activation for the regression task. We add a dropout layer with drop probability $15\%$ between the first and the second hidden layers to avoid overfitting. Each MLP has 227k parameters giving $\sim$450k parameters cumulatively for the complete architecture.

\textbf{\em Training Process:}
We create the training dataset by generating 10k standard compliant transmissions and passing them through a simulated \texttt{wlanTGnChannel} in MATLAB. These are packets distorted by the wireless channel without Additive White Gaussian Noise (AWGN). We collect these packets at the receiver-side, and create a set of time-domain L-LTF OFDM symbols denoted as $X^{(a)}_{\mathrm{train}}$s. We estimate the channel with LS method and save the set of 52 CSI coefficients (one for each subcarrier) as $Y^{(a)}_{\mathrm{train}}$s. Similar to the noise model introduced within the data augmentation pipeline in~\cite{augmentation}, we dynamically add different levels of AWGN to $X^{(a)}_{\mathrm{train}}$s, before feeding them to the NN. This added noise simulates SNRs within the range of 0 to 30 dB with steps of 5 dB. Since $Y^{(a)}_{\mathrm{train}}$s are estimated from noiseless preambles, this dynamic addition of noise helps the NN learn to generalize and associate noisy $X^{(a)}_\mathrm{train}$s with noiseless $Y^{(a)}_\mathrm{train}$s, which improves channel estimator NN performance over its traditional counterpart.

We use Mean Squared Error (MSE) loss function, computed between the output of the MLP and its corresponding ground truth CSI computed through LS. In order to make our model able to process signals received at different power levels, we use Root Mean Square (RMS) normalization on the signal to bring it to a nominal signal power of 1W (i.e. 0 dBW).

\subsection{Demapping} \label{sec:proposed-demapper}
\noindent $\bullet$ \textbf{Classical Approach.} 
After estimating the wireless channel, we equalize the demodulated payload, through dividing the frequency-domain payload by the frequency-domain estimated channel.
The task of the demapper block is to map complex equalized symbols to a sequence of either soft or hard bits in soft-demapping and hard-demapping, respectively. The length of the output bit sequence is proportional to the modulation order. For example, in 16QAM, the transmitter-side \emph{mapper} relates every four bits to one complex symbol. Consequently, in the receiver, the demapper demaps each equalized symbol to four bits.

\noindent $\bullet$ \textbf{Proposed NN Architecture:}
We follow the traditional demapper concept, and propose an NN that generates a sequence of bits for each equalized symbol. Without losing generality, we design an example demapper that demaps symbols from 16QAM. Consequently, our NN has an input size of 2 that represent In-phase (I) and Quadrature (Q) parts of one equalized symbol. Since the demapper NN is supposed to work on one equalized symbol at a time and the symbols are independent from each other, we use a fully-connected NN. Unlike convolutional networks with window size$>$1, our approach operates on each equalized sample separately. The output layer has size 4, where each neuron represents one bit. Since each output bit must be set to either `0' or `1', and multiple bits can be `1' simultaneously, we consider demapping as a multi-label classification problem, and choose \emph{Sigmoid} activation for the last layer.
To determine the number of layers, we search the design space of fully-connected NNs with different number of layers and different number of neurons in each layer. We find out that the smallest NN for the 16QAM demapper is a fully-connected NN with 2 layers, with output sizes of 20 and 4, respectively. This model has only 144 total parameters.

\textbf{\em Training Process:}
The best demapping performance is achieved when the training set contains only low SNR packets. We train the demapper NN with equalized symbols, as $X^{(b)}_\mathrm{train}$s, from 16k packets with SNR 2 dB. For the labels ($Y^{(b)}_\mathrm{train}$s), we use inputs of the mapper at the transmitter-side. In this way, the demapper learns to demap the equalized symbols to the original, non-distorted bits at the transmitter-side.

During inference, the NN receives the equalized symbols. Since we use Sigmoid activation in the last layer that produces outputs in the range of [0,1], we consider the probability of firing each neuron as the probability of corresponding bit being `1'. The soft-bits (i.e., Log Likelihood Ratios (LLRs)) are calculated by taking the logarithm of probability of the output being `0' divided by the probability of the output being `1'. Using this relationship, one LLR value is calculated for each output neuron. These LLR values are scaled by a combination of CSI and noise variance, calculated by the equalizer. Next, the scaled LLRs are de-interleaved and provided to the Forward Error Correction (FEC) decoder to compute the actual transmitted bits, as shown in Fig.~\ref{fig:overview}.

\subsection{Error Correction}\label{sec:proposed-decoder}

\noindent $\bullet$ \textbf{Classical Approach.} 
The role of the forward error correction (FEC) decoder block is to convert the de-interleaved bits to an error-corrected sequence of bits, as shown in Fig.~\ref{fig:overview}. The relationship between the input size and the output size of the decoder depends on the coding rate $C_\mathrm{rate}$, with the output size being $C_\mathrm{rate}$ times the input size. Similar to demapping, conventional decoding can be done in two ways of soft-decoding and hard-decoding, depending on the decoder inputs being whether soft-bits or hard-bits, respectively. In the traditional decoder, the Viterbi~\cite{viterbi1967error} algorithm is used to decode the correct sequence of bits.

\noindent $\bullet$ \textbf{Proposed NN Architecture.}
Unlike the demapper NN where each equalized sample is mapped to a bit sequence independently, decoding from a convolutional code requires processing of a sequence of inputs. For this reason, instead of a simple fully-connected network, the decoder uses Recurrent Neural Networks (RNNs), which are designed to process sequential data. To have the input of the decoder NN between 0 and 1, we convert the de-interleaved LLRs to the probability of a bit being equal to `1'. This probability is achieved by inverting the LLR formula, therefore, probability of each soft-bit (LLR) being `1' equals $1/(1+e^{LLR})$. 

Due to the nature of RNNs, the input size of our proposed decoder NN can vary up to $l$, which corresponds to the size of de-interleaved LLR vector for the payload. The decoder RNN architecture consists of a recurrent part and a fully-connected part stacked together. The recurrent part has 3 Gated Recurrent Unit (GRU) layers, each with 256 units. The fully-connected part has 2 dense layers. The hidden layer has 16 neurons with ReLU activation, and the output layer has 1 neuron. 
Since, same as the demapper, the decoder produces bit sequences, the output layer has Sigmoid activation with size $C_\mathrm{rate}$ times the size of the LLR vector. The decoder NN has $\sim$2.7M total parameters.

\textbf{\em Training Process:}
Similar to demapping, the best decoding performance is achieved when the training set contains only low SNR packets. Therefore, we use 16k packets in SNR 2 dB to train the decoder. As shown in Fig.~\ref{fig:overview}, we train the decoder NN with outputs of de-interleaver at the receiver-side as $X^{(c)}_\mathrm{train}$s and the scrambled bits at the transmitter-side as $Y^{(c)}_\mathrm{train}$s. Using transmitter-side scrambled bits as $Y^{(c)}_\mathrm{train}$s helps the NN learn to map noisy inputs to undistorted original bits, which boosts the decoder performance. 

During inference, the decoder NN generates the probability of each bit being `1'. To yield the final bit sequence, these probabilities are mapped to bit `1' if they are greater than 0.5, and to bit `0' otherwise.

So far, we described how NNs can substitute classical signal processing blocks in the OFDM receiver chain. Next, we describe how the NNs are compressed during training for FPGA deployment.


\subsection{FPGA Implementation}
The NNs described in this paper span a range from small (144 parameters) to large (up to $\sim$2.7M parameters). Direct deployment of such components on resource-constrained hardware (i.e., FPGA) is not possible. To  compress these architectures suitably for FPGA deployment, we use two approaches: 1) Making the network sparser by reducing the number of weights (pruning), and 2) Restricting the weights to be represented by a small number of bits (quantization). The latter not only reduces the memory needed to store each weight, but also fits the limitations of a fixed-point hardware such as FPGA. Moreover, by implementing different quantization methods, we can control which FPGA resources will be used for weight multiplication. Both pruning and quantization methods happen along training and the resulting model has quantized weights where many are set to zero (are pruned).

\begin{figure}[t!!!]
    \centering
    \includegraphics[width=0.9\linewidth]{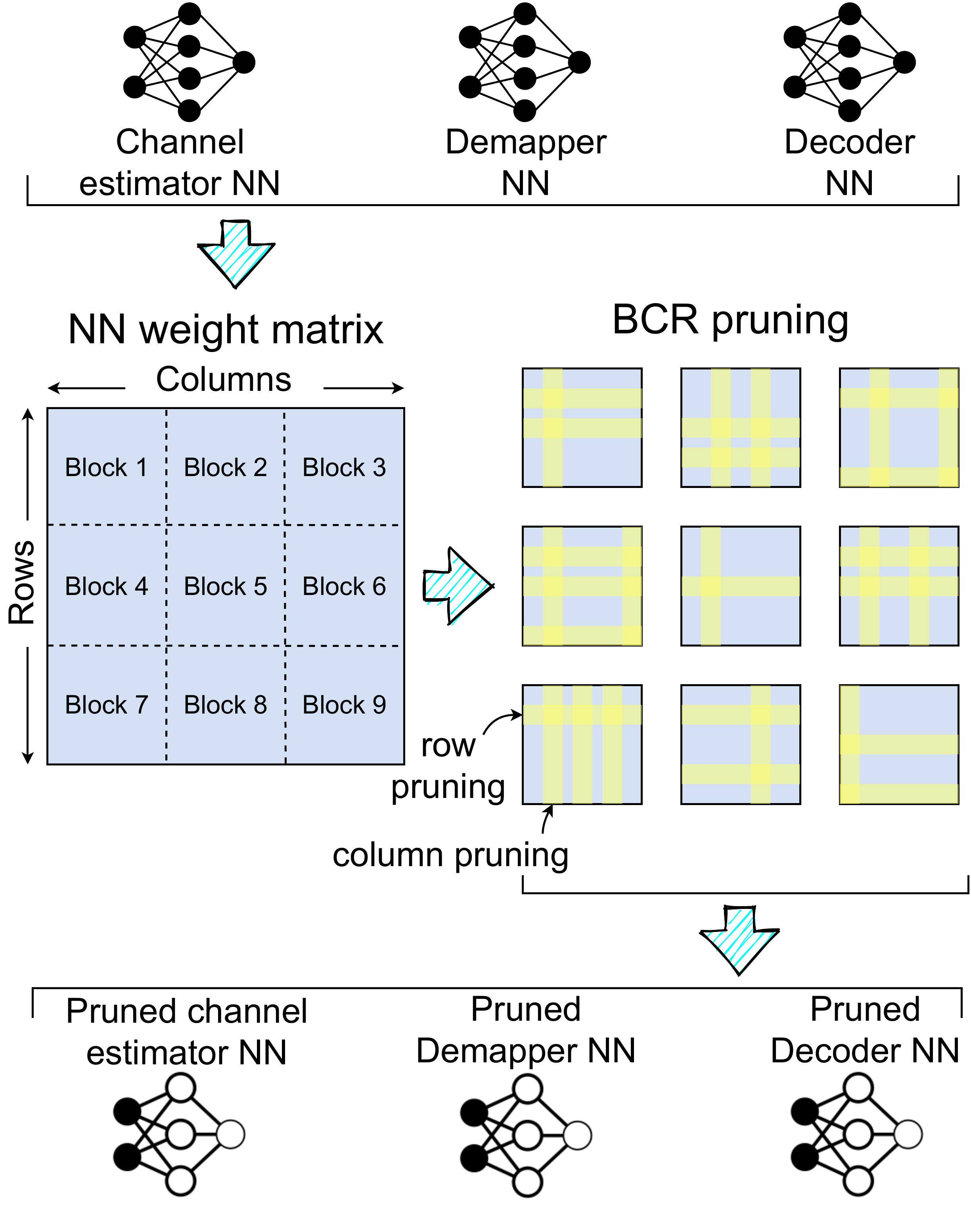}
    \caption{The proposed BCR pruning method is applied to all the fully-connected layers in the channel estimator, demapper, and decoder NNs.}
    \label{fig:pruning}
\end{figure}

\subsubsection{Pruning}
Two popular types of pruning are \textit{structured pruning} and \textit{irregular pruning}. Structured pruning is a coarse-grained pruning approach that removes the whole filter or channel in an NN layer, and is the best for hardware acceleration. However, structured pruning adversely affects the accuracy. In contrast, irregular pruning is a fine-grained pruning approach that sets the weights with small magnitudes to zero, preserves the accuracy, but does not attain acceleration on most hardware platforms. To solve this issue, we propose a Block-based Column Row (BCR) pruning scheme that serves as the universal fine-grained structured pruning. This pruning method can prune convolutional and fully-connected layers. As shown in Fig.~\ref{fig:pruning}, in BCR pruning, each weight matrix is divided into multiple blocks, and row and column pruning is applied to each block separately. We employ ADMM-based pruning~\cite{admm} to determine the row/column pruning ratio automatically. The described fine-grained BCR pruning can significantly outperform the traditional coarse-grained structured pruning, and can provide larger acceleration compared to the more flexible irregular pruning.

\subsubsection{Quantization}
Weight quantization is another method for compressing the NN for FPGA implementation. Two popular methods of quantization are \emph{fixed-point quantization} and \emph{power-of-two quantization}. Fixed-point quantization is a naive quantization method that prepares the NN to run on a fixed-point hardware such as FPGA. Power-of-two quantization converts weight multiplications to simple bit shifts. In this way, weight multiplications, that are typically implemented on the specialized hardware blocks called \emph{DSP48s} inside FPGAs, will be implemented on the \emph{Look-Up Tables (LUTs)} as simple bit shifts. 
Based on these two schemes, we propose a Mixed Scheme Quantization (MSQ) approach that applies fixed-point quantization and power-of-two quantization on different rows of the weight matrix. There are two major motivations for using MSQ: First, different rows of the weight matrix have different distributions. Since power-of-two quantization has higher resolution around the center, it is best to be applied to the rows with a lower variance. In contrast, fixed-point quantization is suitable for rows with near uniform weight distribution, that have higher variance. Second, by using a specific type of quantization for each row, a mix of FPGA resources (DSP48s and LUTs) are used, which balances the resource utilization for NN weight multiplications. We observe that MSQ maintains the accuracy of the two single methods, due to being able to accommodate to different weight distributions.

\section{Performance Evaluation}\label{sec:eval}
We use PyTorch on Nvidia RTX 2080Ti GPUs to train individual NNs.
The metric we use for evaluating the proposed NN-based receiver is the Bit Error Rate (BER), that shows the rate of incorrectly recovered received bits after error correction. 
We evaluate the individual performance of the NNs by inserting each NN in the OFDM receiver, while the rest of the processing is performed by the classical blocks. 
To evaluate the whole NN-based OFDM receiver, we cascade all the 3 NNs and measure the BER. 
For fair comparison, we further create a MATLAB baseline, which demonstrates the BER of a traditional MATLAB OFDM chain, without ML involved, as shown in Fig.~\ref{fig:overview} (bottom). 
In the MATLAB processing chain, for channel estimation, we use the standard compliant LS estimation for preamble-based frequency-domain channel estimation that is implemented in MATLAB function \texttt{wlanLLTFChannelEstimate}. For the demapper, we use soft-demapping through approximate LLR method implemented in MATLAB function \texttt{wlanConstellationDemap}. For the decoder, we use convolutional decoder (that decodes Binary Convolutional Coding (BCC)) implemented in MATLAB function \texttt{wlanBCCDecode}. For BER evaluation, we use simulated and OTA datasets that are briefly described below.

\subsection{Test Dataset Description}
\subsubsection{Simulated dataset} 
We use \texttt{WLAN} toolbox in MATLAB R2020a to create a simulated dataset by generating 192k packets, each containing a random sequence of bits. These packets are then modulated in accordance with IEEE 802.11a standard with Modulation Coding Scheme (MCS) 16QAM 1/2, before passing through a simulated \texttt{wlanTGnChannel} and an AWGN channel with desired SNR level. The SNR levels we use are between 2 dB and 24 dB with steps of 2 dB.
With 192k total packets distributed among 12 SNR levels, we create a test set of 16k packets per SNR.

\subsubsection{Over-The-Air (OTA) dataset}
We collect an OTA dataset in Arena~\cite{arena}, using one transmitter and one receiver. Similar to the simulated data, we generate random bit sequences, modulate them according to IEEE 802.11a standard, and transmit them via Software-Defined Radios (SDRs) placed in an overhead ceiling-mounted array. We repeat this process with different power levels to account for different SNRs, and collect $\sim$17k real-channel-distorted packets with SNRs between 11 to 24 dB, as our OTA test set.

\subsection{Receiver Bit Error Rate Results}

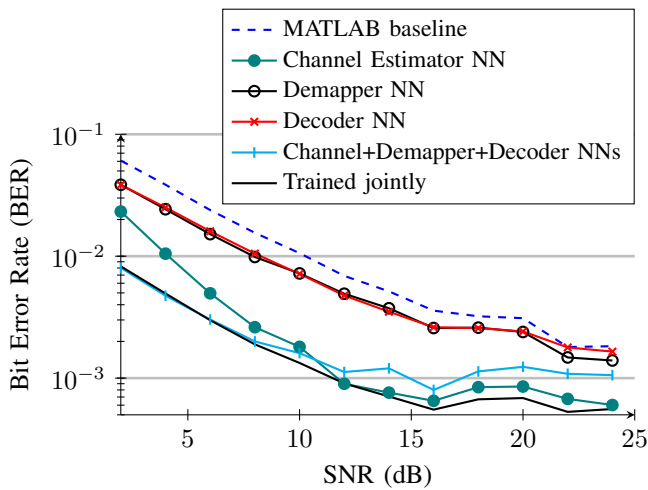
\begin{figure}[t!!!]
\centering
\begin{tikzpicture}
        \begin{axis}[
            axis lines = left,
            xlabel = SNR (dB),
            ylabel = {Bit Error Rate (BER)},
            ymin=0.0005,
            ymax=0.1,
            xmin=2,
            xmax=25,
            ymode=log,
            height=0.6\linewidth,
            width=0.95\linewidth,
            legend style={at={(1,0.75)},
            legend cell align={left},font=\small,
            anchor=south east,legend columns=1},
            ymajorgrids=true,
            grid style={line width=1pt,draw=gray!50}
        ]
        \addplot [thick,color=blue,dashed]
        coordinates{
        (2,0.0608590316772461)(4,0.0385978775024414)(6,0.023909049987793)(8,0.015557975769043)(10,0.0105921020507813)(12,0.00689726257324219)(14,0.00514714050292969)(16,0.00357718658447266)(18,0.00321563720703125)(20,0.00310252380371094)(22,0.00179805755615234)(24,0.00182765197753906)
        };
        \addlegendentry{MATLAB baseline}
        \addplot [thick,color=teal, mark=*]
        coordinates{
        (2,0.0232475433349609)(4,0.0104913482666016)(6,0.00495936584472656)(8,0.00261725616455078)(10,0.00180471038818359)(12,0.000898719787597656)(14,0.000760635375976562)(16,0.000651206970214844)(18,0.000843215942382813)(20,0.000853065490722656)(22,0.000675651550292969)(24,0.000601776123046875)
        };
        \addlegendentry{Channel Estimator NN}
        \addplot [thick,color=black, mark=o]
        coordinates{
        (2,0.0385644455)(4,0.02427315521)(6,0.01516761017)(8,0.009853851318)(10,0.0072239151)(12,0.004922950745)(14,0.003740707397)(16,0.002576026917)(18,0.002594703674)(20,0.002390739441)(22,0.001477722168)(24,0.001395042419)
        };
        \addlegendentry{Demapper NN}
        \addplot [thick,color=red, mark=x]
        coordinates{
        (2,3.8515e-02)(4,0.025105)(6,0.016023)(8,0.010507)(10,0.0071625)(12,0.00476)(14,0.003505)(16,0.002615)(18,0.00257)(20,0.002415)(22,0.001785)(24,0.00165)
        };
        \addlegendentry{Decoder NN}
        \addplot[thick,mark=|, color=cyan]
        coordinates{
        (2,8.0300e-03)(4,4.7400e-03)(6,3.0250e-03)(8,2.0000e-03)(10,1.6025e-03)(12,1.1225e-03)(14,1.2025e-03)(16,8.0000e-04)(18,1.1350e-03)(20,1.2375e-03)(22,1.0850e-03)(24,1.0575e-03)
        };
        \addlegendentry{Channel+Demapper+Decoder NNs}
        
        \addplot[thick, color=black]
        coordinates{
        (2,0.008252687439437984)(4,0.00494332152374031)(6,0.002984352289244186)(8,0.001892131480135659)(10,0.001331592175387597)(12,0.0009049857679263566)(14,0.0007070918120155039)(16,0.0005513717296511628)(18,0.0006700429990310078)(20,0.0006861903464147287)(22,0.0005286079820736434)(24,0.000558116824127907)
        };
        \addlegendentry{Trained jointly}
        \end{axis}
    \end{tikzpicture}
    \caption{BER of simulated dataset achieved from NNs and MATLAB baseline. For the simulated dataset, individual Channel Estimator, Demapper, and Decoder NNs, trained separately, show 86\%, 36\%, and 36\% average BER improvement over MATLAB baseline. The cascade of separately trained NNs provide 61\% BER improvement over MATLAB baseline. If the NNs are trained jointly, BER improvement reaches 77\%, compared to MATLAB baseline.}
    \label{fig:simulated}
\end{figure}

Fig.~\ref{fig:simulated} shows the BER yielded from classical MATLAB processing chain (MATLAB baseline). We also demonstrate the BER when only one of the channel estimator, demapper, or decoder blocks is replaced by the corresponding NN, as well as the end-to-end BER when all the classical blocks are replaced with NNs in the OFDM receiver chain. 

We observe that channel estimator, demapper, and decoder NNs provide up to 86\%, 36\%, and 36\% improvement in BER, respectively, compared to the MATLAB baseline. We observe that the cascade of the 3 NNs trained separately tagged as `Channel+Demapper+Decoder' in Fig.~\ref{fig:simulated} shows 61\% BER improvement compared to MATLAB baseline. We further perform another experiment where we train the 3 NNs jointly. We test the cascade of the trained NNs on the simulated data and show it as `Trained jointly' in Fig.~\ref{fig:simulated}, which improves MATLAB baseline BER by 77\%.

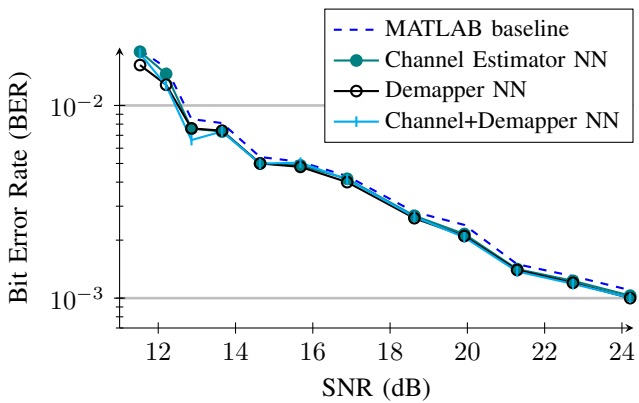
\begin{figure}[t!!!]
\centering
\begin{tikzpicture}
        \begin{axis}[
            axis lines = left,
            xlabel = SNR (dB),
            ylabel = {Bit Error Rate (BER)},
            ymin=0.0007,
            ymax=0.02,
            xmin=11,
            xmax=24.3,
            ymode=log,
            height=0.6\linewidth,
            width=0.95\linewidth,
            legend style={at={(1,0.66)},font=\small,
            legend cell align={left},
            anchor=south east,legend columns=1},
            ymajorgrids=true,
            grid style={line width=1pt,draw=gray!50}
        ]
        \addplot[thick,color=blue,dashed]
        coordinates{
        (11.529,0.0194)(12.2038,0.0157)(12.8607,0.0085)(13.6517,0.0081)(14.6364,0.0054)(15.6802,0.0051)(16.889,0.0043)(18.6287,0.0028)(19.9208,0.0024)(21.2878,0.0015)(22.7319,0.0013)(24.2127,0.0011)
        };
        \addlegendentry{MATLAB baseline}
        \addplot[thick,color=teal, mark=*]
        coordinates{
        (11.529,0.0190094033722438)(12.2038,0.0145740689271818)(12.8607,0.00758567794073631)(13.6517,0.00731967824638927)(14.6364,0.00499830823887668)(15.6802,0.00488979041386629)(16.889,0.00416257413773565)(18.6287,0.00267588116728568)(19.9208,0.00215177973724948)(21.2878,0.00141501613957073)(22.7319,0.00123001129659848)(24.2127,0.00103009879384265)
        };
        \addlegendentry{Channel Estimator NN}
        \addplot[thick,color=black,mark=o]
        coordinates{
        (11.529,0.0162)(12.2038,0.0128)(12.8607,0.0076)(13.6517,0.0074)(14.6364,0.005)(15.6802,0.0048)(16.889,0.004)(18.6287,0.0026)(19.9208,0.0021)(21.2878,0.0014)(22.7319,0.0012)(24.2127,0.001)
        };
        \addlegendentry{Demapper NN}
        \addplot[thick,color=cyan,mark=|]
        coordinates{
        (11.529,0.0185867148415787)(12.2038,0.0126980266814897)(12.8607,0.00661291529482191)(13.6517,0.00733397600244015)(14.6364,0.00498963796311963)(15.6802,0.00504731163777856)(16.889,0.00412904688149297)(18.6287,0.0026423582983257)(19.9208,0.00206875496841136)(21.2878,0.00137564241342658)(22.7319,0.00118604870089119)(24.2127,0.00100323009898904)
        };
        \addlegendentry{Channel+Demapper NN}
        \end{axis}
    \end{tikzpicture}
    \caption{BER of the OTA dataset achieved from NNs and MATLAB baseline. For OTA dataset, Channel Estimator and Demapper NNs provide upto 12\% and 20\% BER improvement over MATLAB baseline. The cascade of these NNs yields an average 10\% BER improvement over all SNRs.}
    \label{fig:real}
\end{figure}

Before testing the OFDM NN-based receiver on the OTA dataset, we need to re-train the channel estimator NN to learn the variations of the real wireless channel. However, the demapper and the decoder NNs do not need to be re-trained. This is because, as shown in Fig.~\ref{fig:overview}, the channel effects are compensated by the equalizer before the data reaches the demapper and the decoder, and hence, demapper and decoder NNs perform independently of the wireless channel. 

As explained in Section~\ref{sec:proposed-channel}, we need data with very high SNR for training the channel estimator NN. Therefore, we collect additional OTA data of totally $\sim$35k packets between SNRs 25 and 37 dB, to re-train the channel estimator NN. Fig.~\ref{fig:real} shows the BER generated by processing the OTA data via the classical MATLAB processing chain (MATLAB baseline), as well as individual and cascaded NN BERs.
For the OTA dataset, the channel estimator and the demapper NNs provide up to 12\% and 20\% BER improvement, respectively, over MATLAB baseline. 
This results in improved BER performance with the end-to-end NN-based receiver showing an average of 10\% BER improvement compared to the MATLAB baseline.

\subsection{FPGA Results}
During the training phase that happens on GPUs, we apply the proposed BCR pruning to all the fully-connected and recurrent layers in the channel estimator, demapper, and decoder NNs. 
Then, we use the proposed MSQ to quantize the NN parameters. 
We run FPGA emulation to estimate FPGA resources for our models, which are shown in Table~\ref{tab:fpga}. 
We verify the speedup introduced by our compression methods on Ettus X310 SDRs with Xilinx Kintex7 T410 FPGA, by measuring the inference latency before and after NN compression. 
As shown in Table~\ref{tab:fpga}, the inference latency of the model decreases by 80\%, 82\%, and 81\% after compression, for the channel estimator, demapper, and decoder NNs, respectively.
We also measure the BER after compression and observe that compression increases the BER averagely by a negligible ratio of 3\%, 1\%, and 4\% in the channel estimator, demapper and decoder NNs, respectively.

\begin{table}[t!!!]
\centering
\resizebox{0.5\textwidth}{!}{%
\begin{tabular}{|l|l:l:l|}
\hline \xrowht[()]{5pt}
NN block & Channel estimator & Demapper & Decoder \\
\hline \xrowht[()]{5pt}
Model type & Dense & Dense & RNN+Dense \\ 
\hline \xrowht[()]{5pt}
Layers & Linear (160, 512) &  Linear (2, 20) & bi-GRU \\
(input size, & Linear (512, 256) & Linear (20, 4) & \big((2,256,3), 512\big) \\
output size) & Linear (256, 52) & & Linear (512, 16) \\
&  & (per-symbol) & Linear (16, 1) \\
\hline \xrowht[()]{5pt}
Pruning Rate & $2.0\times$ & $1.0\times$ & $2.0\times$ \\
\hline \xrowht[()]{5pt}
Weight & \multirow{2}{*}{$8$} & \multirow{2}{*}{$4$} & \multirow{2}{*}{$8$} \\
\xrowht[()]{2pt}
bit-width &&& \\
\hline \xrowht[()]{5pt}
Overall size & \multirow{2}{*}{$8\times$} & \multirow{2}{*}{$8\times$} & \multirow{2}{*}{$8\times$} \\
\xrowht[()]{1pt}
compression &&& \\
\hline \xrowht[()]{5pt}
Working & \multirow{2}{*}{100 MHz} & \multirow{2}{*}{100 MHz} & \multirow{2}{*}{100 MHz} \\
frequency &&& \\
\hline \xrowht[()]{5pt}
Non-compressed & \multirow{2}{*}{$1.67~ms$} & \multirow{2}{*}{$4.97~\mu s$} & \multirow{2}{*}{$210.04~\mu s$} \\
\xrowht[()]{1pt}
latency &&& \\
\hline \xrowht[()]{5pt}
Compressed & \multirow{2}{*}{$0.33~ms$} & \multirow{2}{*}{$0.89~\mu s$} & \multirow{2}{*}{$38.19~\mu s$} \\
latency &&& \\
\hline \xrowht[()]{5pt}
Non-compressed &
\multirow{2}{*}{$4.6\times10^5$} & \multirow{2}{*}{$4.9\times10^5$}& \multirow{2}{*}{$9.08\times10^{10}$} \\
FLOPs &&& \\
\hline \xrowht[()]{5pt}
Compressed &
\multirow{2}{*}{$2.2\times10^5$} & \multirow{2}{*}{$4.9\times10^5$} & \multirow{2}{*}{$4.45\times10^{10}$} \\
FLOPs &&& \\
\hline
\end{tabular}
}
\caption{Compression techniques, overall compression rate and FPGA speedup of the proposed NNs. FLOPs are reported for an L-LTF length of 160 time-domain samples in 5 MHz bandwidth as the Channel estimator NN, packet length of 4128 equalized symbols as the Demapper NN input, which yields 16512 soft-bits for the Decoder NN input.}
\label{tab:fpga}
\end{table}

\subsection{Computational Complexity}
Finally, we compare the computational complexity of the proposed compressed NNs with the traditional counterparts, in terms of FLOPs. 

\noindent \textbf{Channel estimator:} As explained in Section~\ref{sec:proposed-channel} and shown in Fig.~\ref{fig:overview} (bottom), our channel estimator NN replaces both the FFT and the LS algorithms in the standard MATLAB pipeline. We estimate the traditional MATLAB channel estimation to have $\sim$4.4 $\times$ $10^3$ FLOPs for our L-LTF length of 160 time-domain complex samples. 
By comparing this value to the compressed channel estimator NN FLOPs shown in the last row of Table~\ref{tab:fpga}, we observe that NN FLOPs are $\sim$50 $\times$ the traditional algorithm.
    
\noindent \textbf{Demapper:} The traditional demapping algorithm that we use in MATLAB is implemented through \emph{approximate LLR} method. We estimate the complexity of this algorithm to be 82 FLOPs per generated soft-bits for 16QAM, as this algorithm calculates the distance of each equalized symbol from all the known symbols in the constellation. The FLOPs add up to $\sim$1.3 $\times$ $10^6$ for our packet length of 4128 equalized 16QAM symbols (4128$\times$4 soft-bit). By comparing this value to the compressed demapper NN FLOPs shown in the last row of Table~\ref{tab:fpga}, we observe that NN FLOPs are $\sim$0.37 $\times$ those of the traditional algorithm.
    
\noindent \textbf{Decoder:} The traditional decoding algorithm that we use in MATLAB is implemented through Viterbi algorithm, as explained in Section~\ref{sec:proposed-decoder}. We calculate the decoder function complexity to be $\sim$5.6 $\times$ $10^8$ FLOPs for our decoder input length of 16512 soft-bits. 
By comparing traditional algorithm FLOPs to the compressed decoder NN FLOPs shown in the last row of Table~\ref{tab:fpga}, we observe that NN FLOPs are $\sim$79$\times$ the traditional algorithm.

The total FLOPs for the three compressed NN blocks add up to 4.45 $\times$ $10^{10}$ which is compared with total FLOPs for the traditional algorithms, 5.61 $\times$ $10^8$. We observe that FLOPs count of the proposed compressed NNs are overall 79 $\times$ the cumulative FLOPs for the traditional algorithms. This opens up new research topics to study the tradeoff and identify switching instances, as discussed in Section~\ref{sec:future}.



\section{Open Research Challenges}\label{sec:future}
\begin{enumerate}
    \item \textbf{Processing granularity:} 
    In our proposed NN-based scheme, there is a difference between processing granularity for different NNs. The channel estimator processes one packet at a time, however, the demapper granularity is one equalized sample. This granularity gap opens up opportunities for ways to parallelize the operations via an ensemble of demapper NNs that demap successive samples in parallel. This brings up interesting resource-performance planning and tradeoffs in the choice of FPGA size versus the possible speedup in time.
    \item \textbf{Performance under different environments or configurations:} Our results show that properly compressed NN-based receiver provides better BER performance and has fewer FLOPs compared to the traditional MATLAB receiver. However, the NN-based receiver has its limitations. For example, different channel estimators, demappers, and decoders need to be trained for different environments, different modulation schemes, and different coding rates, respectively. These further impose larger memory requirements to store weights for multiple NNs for each block. Methods such as transfer learning and life-long learning with pruning~\cite{learn-prune-share} can be explored to reuse and share a portion of NN weights among different configurations, and reduce large memory requirements.
    \item \textbf{Identifying switching instances between classical and NN blocks:} We have shown that the NN-based OFDM receiver provides better performance compared to the classical one, in a variety of circumstances. However, a purer NN-based receiver can consume more resources and power compared to the traditional receiver. Since our NN-based receiver is modular, the logic that determines which modules to introduce into the receiver chain and when, is a completely new area of research. As an example, this decision may be made at run-time, based on desired reception performance and available on-board resources.
\end{enumerate}



\section{Conclusion}\label{sec:con}
In this paper, we proposed a model-driven design for NN-based OFDM receivers. Our receiver chain consists of 3 NNs for channel estimation, symbol to bit demapping, and error correction decoding. The NNs were designed based on wireless domain knowledge, and trained independently with data acquired from different data parts in the traditional transmitter and receiver. The trained networks were then cascaded to compose the complete receiver chain. The proposed NN-based receiver was evaluated with both simulated and OTA datasets, and showed averagely 61\% and 10\% improvement in BER compared to the traditional solution, when tested with simulated and OTA datasets, respectively. We further proposed two methods of pruning and quantization to compress our NNs and prepare them for FPGA implementation. We also showed that despite the BER performance gain, the proposed compressed NNs FLOPs are $\sim$79$\times$ their traditional counterparts. This complexity-performance trade-off opens up new research opportunities as discussed in Section~\ref{sec:future}.


\section*{Acknowledgement}
This work is supported by DARPA SPiNN HR00112090055, DARPA LwLL SC1821301, NSF 1923789, and NSF 1845833 awards.

\bibliographystyle{ieeetr} 
\bibliography{ref} 


\section*{Biographies}

\vskip -3\baselineskip plus -1fil
\begin{IEEEbiographynophoto}
{Nasim Soltani} is a Ph.D. candidate at the Institute for Wireless IoT at Northeastern University, advised by professor Chowdhury. Her area of interest is AI-aided algorithms for applications in the physical layer of wireless communications systems.
\end{IEEEbiographynophoto}

\vskip -2\baselineskip plus -1fil
\begin{IEEEbiographynophoto}
{Hai Cheng} is a Ph.D. candidate in Computer Engineering at the Institute for Wireless IoT at Northeastern University. He received his B.Eng degree in 2015 from Xidian University, China, and master degree in 2018 from ShanghaiTech University, China. His research interests include machine learning and optimization in wireless network systems.
\end{IEEEbiographynophoto} 

\vskip -2\baselineskip plus -1fil
\begin{IEEEbiographynophoto}
{Mauro Belgiovine} is pursuing his Ph.D. at the Electrical and Computer Engineering department at Northeastern University, under the guidance of Professor Kaushik Chowdhury. His current research interests involve deep learning, wireless communications, and heterogeneous computing.
\end{IEEEbiographynophoto}

\vskip -2\baselineskip plus -1fil
\begin{IEEEbiographynophoto}
{Yanyu Li} is a Ph.D. candidate at the Department of Electrical and Computer Engineering in Northeastern University, advised by Professor Yanzhi Wang. His research interests include deep learning, neural network architecture search, pruning and quantization. 
\end{IEEEbiographynophoto} 

\vskip -2\baselineskip plus -1fil
\begin{IEEEbiographynophoto}
{Haoqing Li} is a Ph.D. candidate in Electrical and Computer Engineering at Northeastern University, Boston, MA. got his BS degree in Electrical Engineering from Wuhan University, China and MS degree in Electrical and Computer Engineering at Northeastern University, Boston, MA. His research interests include GNSS signal processing, anti-jamming technology and robust statistics.
\end{IEEEbiographynophoto}

\vskip -2\baselineskip plus -1fil
\begin{IEEEbiographynophoto}
{Bahar Azari} is a Ph.D. student at the center for signal processing, imaging, reasoning, and learning (SPIRAL) of Northeastern University. She received her B.Sc. in electrical engineering from the Amirkabir University of Technology in 2011 and her M.Sc. in telecommunications engineering from Politecnico di Milano in 2014. Her research interests include applied signal processing and machine learning with expertise in deep generative models and latent variable models.
\end{IEEEbiographynophoto}

\vskip -2\baselineskip plus -1fil
\begin{IEEEbiographynophoto}
{Salvatore D'Oro} is a Research Assistant Professor with the Institute for the Wireless IoT at Northeastern University, USA. He received his Ph.D. from the University of Catania in 2015. He serves on the technical program committee of IEEE INFOCOM and the Elsevier Computer Communications journal.
His research interests include optimization and learning in NextG systems.
\end{IEEEbiographynophoto}

\vskip -2\baselineskip plus -1fil
\begin{IEEEbiographynophoto}
{Tales Imbiriba} received his Doctorate degree from the Department of Electrical Engineering (DEE) of the Federal University of Santa Catarina (UFSC), Florian\'opolis, Brazil, in 2016. He served as a Postdoctoral Researcher at the DEE--UFSC and is currently a Postdoctoral Researcher at the ECE dept. of the Northeastern University, Boston, MA, USA. 
His research interests include audio and image processing, pattern recognition, kernel methods, adaptive filtering, and Bayesian Inference.
\end{IEEEbiographynophoto}

\vskip -2\baselineskip plus -1fil
\begin{IEEEbiographynophoto}
{Tommaso Melodia} is a Professor at Northeastern University. He has been named William Lincoln Smith Professor in recognition of his significant research contributions and exceptional leadership in the field of electrical and computer engineering. He is the Director of the Institute for the Wireless IoT, and the Director of Research for the PAWR Project Office. He received his Ph.D. degree in Electrical and Computer Engineering from Georgia Institute of Technology in 2007. 
His research focuses on modeling, optimization, and experimental evaluation of wireless networked systems. He serves as Editor-in-Chief for Computer Networks.
\end{IEEEbiographynophoto}
 
\vskip -2\baselineskip plus -1fil
\begin{IEEEbiographynophoto}
{Pau Closas} is an Assistant Professor at Northeastern University,  Boston,  MA. He received the MS and Ph.D. degrees in Electrical Engineering from UPC in 2003 and 2009.  He also holds a MS in Advanced Mathematics from UPC, 2014.  His primary areas of interest include statistical signal processing, robust stochastic filtering, and machine learning, with applications to positioning systems and wireless communications.  He is the recipient of the 2014 EURASIP Best Ph.D. Thesis Award, the 9th Duran Farell Award, the 2016 ION Early Achievements Award, and a 2019 NSF CAREER Award.
\end{IEEEbiographynophoto}

\vskip -2\baselineskip plus -1fil
\begin{IEEEbiographynophoto}
{Yanzhi Wang} is currently an Assistant Professor at the Department of ECE at Northeastern University, Boston, MA. 
His research focuses on model compression and platform-specific acceleration of deep learning architectures, maintaining the highest model compression rates on representative DNNs.
He received the U.S. Army Young Investigator Program Award (YIP), Massachusetts Acorn Innovation Award, Ming Hsieh Scholar Award, and other research awards from Google, MathWorks, etc. His recent research achievement, CoCoPIE, can achieve real-time performance on almost all deep learning applications using off-the-shelf mobile devices, outperforming competing frameworks by up to 180X acceleration.
\end{IEEEbiographynophoto}

\vskip -2\baselineskip plus -1fil
\begin{IEEEbiographynophoto}
{Deniz Erdogmus}(Senior Member, IEEE) is a Professor of ECE at Northeastern University, Boston, MA. He received his Ph.D. degree in electrical and computer engineering from the University of Florida, Gainesville, FL, in 2002. He held a postdoctoral position at the University of Florida, until 2004. His researches focus on statistical signal processing and machine learning with applications to contextual signal/image/data analysis with applications in cyber-human systems. 
\end{IEEEbiographynophoto}

\vskip -2\baselineskip plus -1fil
\begin{IEEEbiographynophoto}
{Kaushik Chowdhury} (M’09-SM’15) is a Professor at Northeastern University, Boston, MA. He received his Ph.D. degree from Georgia Institute of Technology in 2009. His current research interests involve systems aspects of networked robotics, machine learning for agile spectrum sensing/access, wireless energy transfer, and large-scale experimental deployment of emerging wireless technologies.
\end{IEEEbiographynophoto}

\end{document}